\documentclass[aps,pra,reprint,amsmath,amssymb]{revtex4-2}

\usepackage{amsmath}
\usepackage{tipa}
\usepackage{bbm}
\usepackage{txfonts}
\usepackage{graphicx}
\usepackage{dcolumn}
\usepackage{comment}
\usepackage{bm}
\usepackage[mathscr]{eucal}
\usepackage{amssymb}
\usepackage{amsfonts}
\usepackage{latexsym}
\usepackage{color}
\usepackage{graphicx}
\usepackage{xfrac}
\usepackage{mathrsfs} 
\usepackage[colorlinks=true,linkcolor=blue,citecolor=blue,urlcolor=blue,]{hyperref}
\usepackage[doipre={doi:~}]{uri}

\begin{document}


\title{Optimized Slice-Phase Control of Mirror Pulse in Cold-Atom Interferometry with Finite Response Time}
\author{Xueting Fang$^{1,2}$}
\author{Doudou Wang$^{1}$}
\author{Kun Yuan$^{1}$}
\author{Jie Deng$^{1}$}
\author{Qin Luo$^{1}$}\email[E-mail: ]{luoqin@hust.edu.cn}
\author{Xiaochun Duan$^{1}$}
\author{Minkang Zhou$^{1}$}
\author{Lushuai Cao$^{1}$}
\author{Zhongkun Hu$^{1,3}$}

\affiliation{$^1$National Gravitation Laboratory, MOE Key Laboratory of Fundamental Physical Quantities Measurement, and School of Physics, Huazhong University of Science and Technology, Wuhan 430074, People's Republic of China\\
$^2$Wuhan Gravitation and Solid Earth Tides National Observation and Research Station, Wuhan Hubei, 430071\\
$^3$Wuhan Institute of Quantum Technology, Wuhan 430206, People’s Republic of China} 

\date{\today}

\begin{abstract}
Atom interferometers require both high efficiency and robust performance in their mirror pulses under experimental inhomogeneities. In this work, we demonstrated that quantum optimal control designed mirror pulse significantly enhance interferometer performance by using novel adaptive sliced structure. Using gradient ascent pulse engineering (GRAPE), optimized mirror pulse for a Mach-Zehnder light-pulse atom interferometer was designed by discretizing the control into non-uniform phase slices. This design broadened the tolerence to experimentally relevant variations in detuning $[-\Omega_0,\Omega_0]$ and Rabi frequency $[0.1\times\Omega_0,1.9\times\Omega_0]$ ($\Omega_0=2\pi\times25$ kHz), while maintaining high transfer efficiency even when the response-time delays up to 1.6 $\rm{\mu s}$. The optimized pulse was found to be robust to coupling inhomogeneity and velocity spread, offering a significant improvement in robustness over conventional pulse. The adaptive pulse slicing method provides a minimalist strategy that reduces experimental complexity while enhancing robustness and scalability, offering an innovative scheme for quantum optimal control in high precision atom interferometry.
\end{abstract}


\maketitle

\section{Introduction}
Atom interferometers (AIs), as an important tool for precision measurement, have been widely used in measuring gravitational acceleration \cite{Peters1999,LouchetChauvet2011,Poli2011,Hu2013,Wu2019,Zhang2021}, gravity gradient \cite{Snadden1998,Duan2014,Caldani2019,Janvier2022,Stray2022}, rotation \cite{Gustavson1997,Berg2015,Xu2020,Yao2021,Gautier2022}, Newton’s gravitational constant \cite{Fixler2007,Rosi2014,Mao2021}, fine structure constant \cite{Wicht2002,Bouchendira2011,Parker2018}, as well as in testing equivalence principle \cite{Duan2016,Rosi2017,Zhang2020,Wang2023,Li2023}, detecting dark matter and dark energy \cite{Hamilton2015,Burrage2016,Sabulsky2019}, and have also been proposed to detect gravitational waves \cite{Dimopoulos2009,Yu2011}. An essential way to improve the measurement precision of these physical quantities is to enhance the precision of AIs. Therefore, how to improve the precision of AIs is an important issue.

The ultimate sensitivity of AIs is fundamentally constrained by three key parameters: the de Broglie wavelength of the atomic matter wave, the interferometric area, and the number of atoms. Research groups worldwide were pursuing various strategies to enhance interferometer sensitivity, including the use of ultracold atoms \cite{Grond2010,Berrada2013,Zhang2016} and large-scale fountain configurations with baselines up to 10 meters \cite{Dickerson2013,Zhan2020,Zhao2022}. In recent years, inspired by major advances in pulse optimization techniques from nuclear magnetic resonance (NMR) spectroscopy \cite{Levitt1979,Baum1985,Dunning2014}, there has been growing interest in applying quantum optimal control to improve the precision of AIs.

The preparation of optimized pulses is critical for enhancing AI’s performance, as these pulses can substantially improve both the reflectivity of $\pi$ pulse and the robustness to variations in laser detuning and Rabi frequency, thereby increasing fringe contrast and measurement sensitivity. In particular, tailored error-robust pulses have been extensively investigated through theoretical proposals \cite{Saywell2018,Saywell2020a,Louie2023,Chen2023}, which demonstrated that quantum control techniques can improve interferometer robustness and performance. By contrast, experimental realizations remain limited, with a few notable demonstrations \cite{Mueller2008,Saywell2023,Saywell2020} employing composite Floquet pulses, customized Raman pulses, or numerically optimized pulses to enhance fidelity and fringe contrast. Despite these advances, these pulses often exhibit complex temporal phase profiles, whose experimental realization poses considerable challenges due to the finite response of phase modulators. Moreover, the effect of the modulator's response time on the achieved fidelity remains largely unexplored in experimental implementations. 

To overcome these limitations, we introduce a slice phase optimization strategy that incorporates the spatially density and velocity distributions of the atomic ensemble. This approach generates optimized pulses exhibiting a step-like phase profile, implemented by discretizing the laser phase difference into finite-duration time slices with constant phase values. The proposed method significantly enhances Raman transfer efficiency and mitigates signal loss caused by transverse cloud expansion and longitudinal velocity dispersion, while simultaneously reducing waveform complexity. Moreover, the finite response time in experiment is explicitly accounted for, enhancing experimental feasibility without sacrificing high-fidelity performance.

This paper is organized as follows: In Sec.\ref{II}, we describe the theoretical model of the system, including the spatial and velocity distributions of the atomic cloud and the internal degrees of freedom, and present the system Hamiltonian. In Sec.\ref{III}, the optimal control framework is introduced, including the configuration of key optimization parameters, and cover both single-particle fidelity and ensemble-averaged fidelity. In Sec.\ref{IV}, numerical results achieved with the optimized pulses are presented, including those that account for response time, demonstrating significant fidelity enhancement. Finally, a conclusion is given in Sec.\ref{V}.

\section{Theoretical Model}\label{II}
{In this work,  we model an atomic ensemble consisting of $10^6$ $^{87}\rm{Rb}$ atoms, whose spatial and velocity distributions are assumed to follow typical Gaussian profiles used in AI simulations. Since the Raman beams propagate along the $z$-axis, only the longitudinal velocity component $v_z$ and the transverse spatial distribution in the $x-y$ plane are relevant for the effective detuning and coupling strength experienced by the atoms. The ensemble parameters—corresponding to a transverse temperature of about 3 $\mu$K in the $x-y$ plane and an effective longitudinal temperature below 300 $\rm{nK}$—are introduced as representative values commonly encountered in experiments after moving molasses and Raman velocity selection, and are used to construct the distribution functions for numerical simulation.}

\begin{figure}[t]
\includegraphics[trim=40 5 40 5,width=\columnwidth]{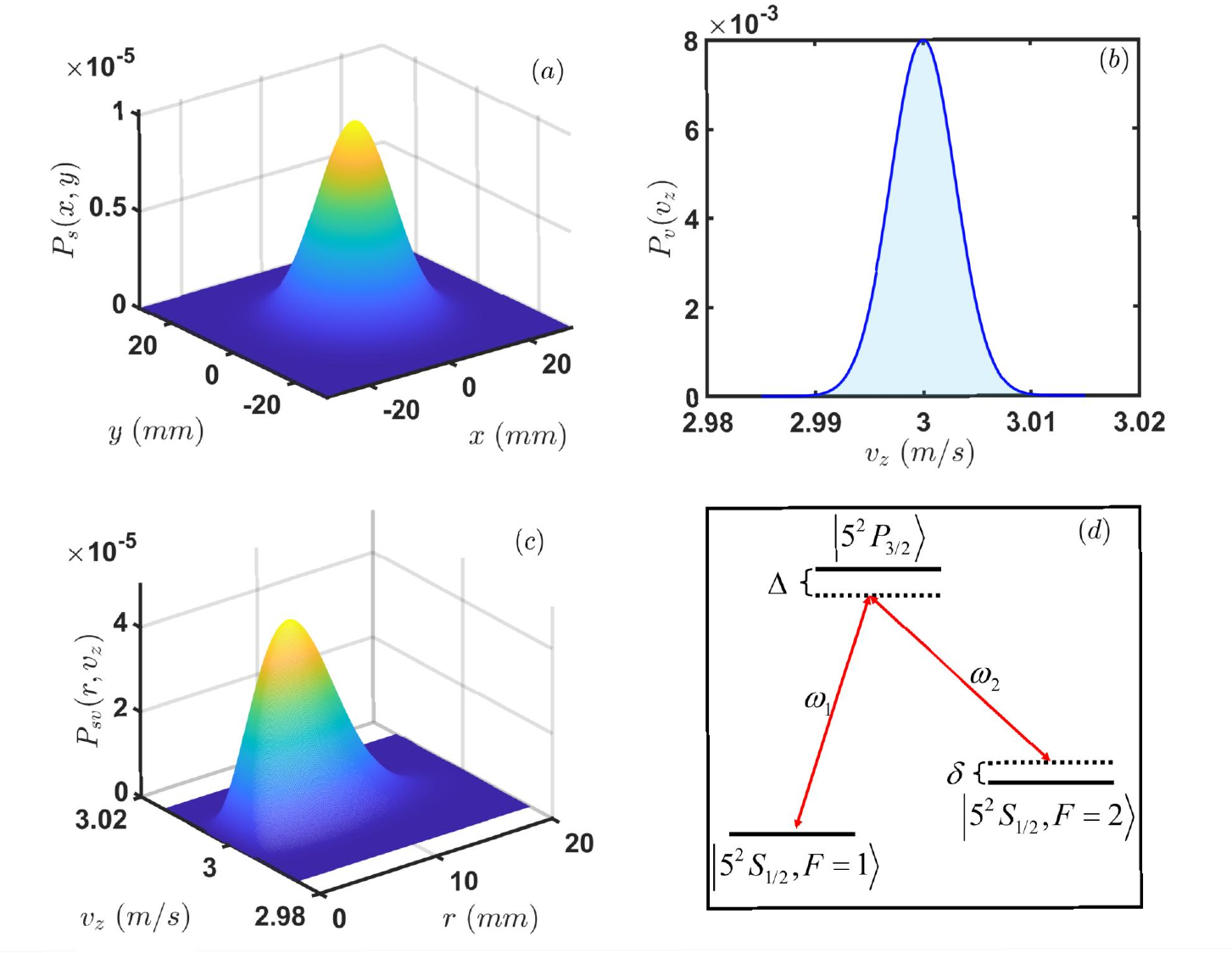}
\caption{\label{fig-pr}  {(a) Spatial distribution of the atomic cloud in the $x-y$ plane, assumed as an ideal Gaussian profile in our numerical model. (b) Velocity distribution of atoms in the $z$-direction follows a Gaussian profile. (c) The combined probability distribution constructed from the assumed spatial and velocity Gaussian profiles.} (d) Schematic of the three-level Raman system. Two counter-propagating laser beams with frequencies $\omega_1$  and $\omega_2$ couple the two hyperfine ground states $|g\rangle$ and $|e\rangle$ via the intermediate excited state $|i\rangle$. The single-photon detuning $\Delta$ and the two-photon detuning $\delta$ are labeled.}
\end{figure}

As shown in Fig. \ref{fig-pr}(a), the spatial density distribution in the transverse x-y plane is assumed as
\begin{equation}
\begin{aligned}
{P_s}(x,y) = \frac{1}{{2\pi {\sigma _x}{\sigma _y}}}\exp \left( { - \frac{{{x^2}}}{{2\sigma _x^2}} - \frac{{{y^2}}}{{2\sigma _y^2}}} \right),
\label{P_r}
\end{aligned}
\end{equation}
where $\sigma_x$ and $\sigma_y$ represent the position broadening of the atomic cloud in the $x$ and $y$ directions, respectively. To simplify the description and to exploit the cylindrical symmetry of the system, the Cartesian coordinates $(x, y)$ were transformed into cylindrical coordinates $(r,\theta)$, where $r = \sqrt {{x^2} + {y^2}} $ and $\theta  = {\tan ^{ - 1}}\left(y/x\right)$. Under the assumption of azimuthal symmetry in both the trapping potential and laser configuration, the spatial density reduces effectively to a function of only the radial coordinate $r$ , i.e., ${P_s}\left( r \right) = ({r/ {\sigma _{\rm{r}}^2}})\exp \left( { - {{{r^2}} / {2\sigma _{\rm{r}}^2}}} \right)$ with ${\sigma _r} = {\sigma _x} = {\sigma _y}$. Similarly, the atomic velocity distribution along the longitudinal $z$-axis followed a Gaussian form, as shown in Fig  \ref{fig-pr}(b) , and is assumed as
\begin{equation}
\begin{aligned}
{P_v}\left( {{v_z}} \right) = \frac{1}{{\sqrt {2\pi \sigma _v^2} }}\exp \left( { - \frac{{v_z^2}}{{2\sigma _v^2}}} \right), \label{P_v}
\end{aligned}
\end{equation}
where $\sigma _v = \sqrt{{{k_B}T}/M}$ represents the velocity broadening along $z$-direction, with $k_B$ being the Boltzmann constant, $T$ the temperature, and $M$ the atomic mass. Furthermore, to account for possible correlations between the transverse spatial distribution and the longitudinal velocity distribution, a combined probability function ${P_{sv}}\left( {r,{v_z}} \right) = {P_s}\left( r \right) \cdot {P_v}\left( {{v_z}} \right)$ was introduced and numerically characterized. The form of this combined distribution is illustrated in Fig. \ref{fig-pr}(c).

In this model, the effect of transverse thermal expansion was omitted. At a transverse temperature of 3 $\mu$K, the thermal velocity was approximately 1.7 cm/s, corresponding to a transverse displacement on the order of a few micrometers over a Raman pulse duration of  $\sim \mu \rm{s}$. This displacement was negligible compared to the typical laser beam waist, ensuring that the laser intensity experienced by each atom remained effectively constant. Therefore, the influence of transverse motion on the transition dynamics was safely ignored.

Coherent manipulation of the internal states of the atomic ensemble was realized by employing a Raman-type AI based on the D2 line of $^{87}$Rb. The two hyperfine ground states, denoted as $\left| g \right\rangle  = \left| {{5^2}{S_{1/2}},F = 1} \right\rangle $ and $\left| e \right\rangle  = \left| {{5^2}{S_{1/2}},F = 2} \right\rangle $, were coupled via two-photon stimulated Raman transition. An intermediate excited state  $\left| i \right\rangle  = \left| {{5^2}{P_{3/2}}} \right\rangle $ was introduced to enable a two-photon Raman process, in which a photon is absorbed from one laser beam and emitted into the other, thereby producing a coherent population transfer between $|g\rangle$ and $|e\rangle$ . The transition is accompanied by a momentum transfer $\hbar {{\bf{k}}_{{\rm{eff}}}} = \hbar \left( {{{\bf{k}}_1} - {{\bf{k}}_2}} \right)$, where $\bf{k_1}$ and $\bf{k_2}$ are the wave vectors of the two laser fields. 

The energy level configuration and the Raman coupling scheme are illustrated in Fig. \ref{fig-pr}(d). Each Raman beam was detuned from the intermediate state $|i\rangle$ by a large single-photon detuning $\Delta$, defined as $\Delta=(\omega_i-\omega_g)-\omega_1$, which is the difference between the laser frequency $\omega_1$ and the transition frequency from $|g\rangle$ to $|i\rangle$. This detuning allows the population in $|i\rangle$ to be adiabatically eliminated. The two-photon detuning, defined as $\delta  = \left( {{\omega _1} - {\omega _2}} \right) - {\omega _{eg}}$, quantifies the frequency mismatch between the frequency difference of the two lasers $\left( {{\omega _1} - {\omega _2}} \right)$ and the ground state hyperfine splitting $ {\omega _{eg}}=\omega_e-\omega_g$. Both $\Delta$ and $\delta$ play key roles in determining the Rabi frequency and the fidelity of the state transfer.

In practice, the Raman beams exhibit a finite transverse profile, which induces a position dependent variation in the local Rabi frequency. To account for this spatial inhomogeneity, each beam was modeled with a Gaussian transverse profile and assumed to counter-propagate along the $z$-axis, with the electric field at a transverse position $r$ expressed as 
\begin{equation}
\begin{aligned}
{E_j}\left( {r,t} \right) = {E_{0j}}\exp \left( { - \frac{{{r^2}}}{{{w^2}}}} \right)\exp \left[ {i\left( {{k_z}_jz - {\omega _j}t + {\phi _j}} \right)} \right],\left( {j = 1,2} \right), \label{Exy}
\end{aligned}
\end{equation}
where $ {E_{0j}}$ is the peak field amplitude of the $j$-th Raman laser beam, $w$ is the laser beam waist, $\phi_j$ is the laser phase, and $k_{zj}$ is the wavevector along $z$. Consequently, atoms at different radial positions experience different field intensities, leading to varying Rabi frequencies across the atomic cloud, thereby introducing spatial dependence into the atomic dynamical evolution.

The atomic ensemble was initialized by randomly sampling positions $r$ and longitudinal velocities $v_z$ from Gaussian distributions with measured widths $\sigma_r$ and $\sigma_{v_z}$. To capture this behavior, we sampled the atoms to obtain the basis states:
\begin{equation}
\begin{aligned}
{\left| g \right\rangle _{r,{v_z}}} \equiv \left| {g;r,{v_z}} \right\rangle ,\\
{\left| e \right\rangle _{r,{v_z}}} \equiv \left| {e;r,{v_z}} \right\rangle, \label{ge}
\end{aligned}
\end{equation}
Each atom was prepared in the ground state $|g\rangle$,  and the initial state of a single atom was represented as $\left| {\psi \left( 0 \right)} \right\rangle  \equiv {\left| g \right\rangle _{r,{v_z}}}$.

Under the rotating wave approximation and in the limit of large single-photon detuning, the system reduces to an efficient two-level model. The effective Hamiltonian of a two-level atom interacting with the laser fields can be expressed as $\hat H\left( {r,{v_z},t} \right) = {\hat H_0}\left( {{v_z}} \right) + {\hat H_{{\rm{control}}}}\left( {r,t} \right)$, where
\begin{equation}
\begin{aligned}
{{\hat H}_0}\left( {{v_z}} \right) &= \frac{{\hbar \delta_{\rm{eff}} \left( {{v_z}} \right)}}{2}{{\hat \sigma }_z},\\ 
{{\hat H}_{{\rm{control}}}}\left( {r,t} \right) &= \frac{{\hbar {\Omega _R}\left( {r} \right)}}{2}\left\{ {\cos \left[ {{\phi _L}\left( t \right)} \right]{{\hat \sigma }_x} + \sin \left[ {{\phi _L}\left( t \right)} \right]{{\hat \sigma }_y}} \right\}. \label{H_con}
\end{aligned}
\end{equation}
Here $\hat{\sigma_x}$, $\hat{\sigma_y}$ and $\hat{\sigma_z}$ are the Pauli 
matrices, ${\phi _L}\left( t \right)=\phi_1(t)-\phi_2(t)$ is the relative laser phase, i.e., the phase difference between the two Raman beams, $\delta(v_z)$ is the effective two-photon detuning, and $\Omega_R(r)$ is the two-photon Rabi frequency. The detuning can be further decomposed as $\delta_{\rm{eff}} \left( {v_z} \right) =  - {\delta ^{\rm{AC}}} + \left( {\omega _1} - {\omega _2} \right) - \left( {\omega _{eg} + \delta _{\rm{rec}} + \delta _D} \right)$, where $\delta^{\rm{AC}}$ represents the ac Stark shift, $\delta _{\rm{rec}} = {{\hbar {{\left| {{{\bf{k}}_{{\rm{eff}}}}} \right|}^2}}/{2M}} $ is the recoil shift term, and ${\delta _{\rm{D}}} = {k_{{\rm{eff}}}{v_z}}$ is the Doppler shift from the longitudinal atomic velocity $v_z$. {In the following, we use $\delta_{\rm{eff}}$ as the detuning parameter, since it incorporates the Doppler shift together with the additional light-shift corrections. For clarity, we write simply “detuning” to refer to $\delta_{\rm{eff}}$ throughout the rest of the paper.} The Rabi frequency $\Omega_R(r)$ can be expressed as ${\Omega _R}\left( r \right) = {{{\Omega _1}\left( r \right){\Omega _2}\left( r \right)}}/{{2\Delta }}$, where $\Omega_1(r)$ and $\Omega_2(r)$ represent the single-photon Rabi frequencies, which characterize the strength of the dipole coupling between each laser field and the atomic transition. 

By treating the relative laser phase ${\phi _L}\left( t \right)$ as a tunable control parameter, the effective Hamiltonian provides a unified and systematic framework for pulse shaping and optimal control. In practice, a tailored phase modulation strategy can be implemented to improve robustness against inhomogeneous broadening.

\section{OPTIMAL CONTROL}\label{III}
\begin{figure}[t]
\includegraphics[trim=0 5 0 5,width=\columnwidth]{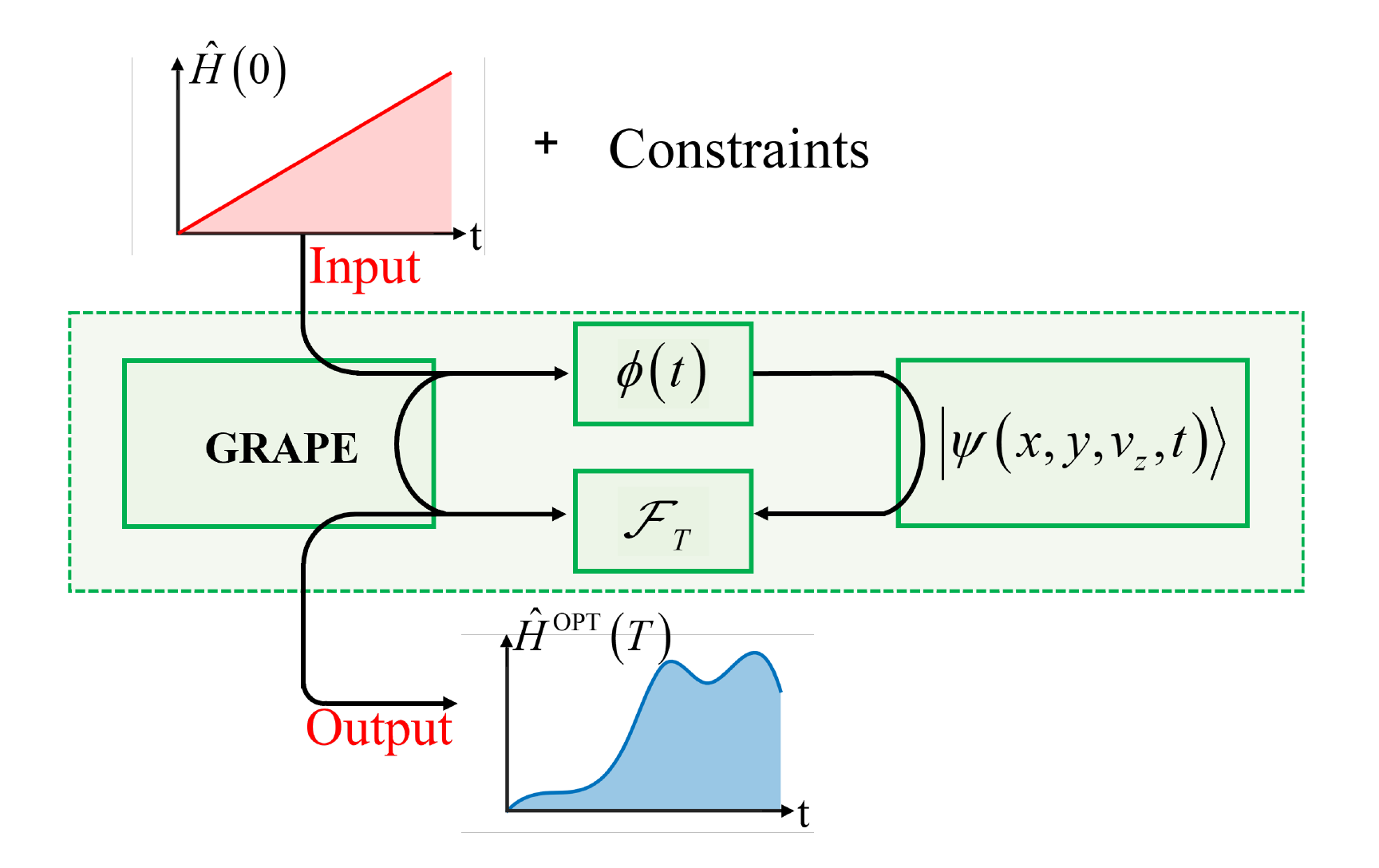 }
\caption{\label{fig-optimal} Sketch of the GRAPE optimal control algorithm. An initial control field $\hat{H}(0)$ is applied to a numerically simulated system and iteratively optimized under physical constraints. The optimization maximizes a figure of merit $\cal{F}$ after time evolution, yielding the final control field $\hat{H}^{\rm{OPT}}(T)$. }
\end{figure}

Optimal control theory provides a systematic framework for designing time-dependent control fields that maximize a chosen performance metric at the end of the system’s evolution. As shown in Fig. \ref{fig-optimal}, the system dynamics were governed by the control Hamiltonian {$\hat H\left( t \right) = {\hat H_0} + {\hat H_{{\rm{control}}}}\left( {{\phi(t)}} \right)$}, where $\hat{H}_0$ denotes the free Hamiltonian and ${\hat H_{{\rm{control}}}}$ represents the external control fields parameterized by {$\phi(t)$}. In our work, {$\phi(t)$} corresponded to the time-dependent modulation of the Raman laser phase, and its optimization relies on the performance function $\cal{F}$ following the time evolution equation ${{i\hbar d\left| {\psi \left( t \right)} \right\rangle }/ {dt}} = \hat H\left( t \right)\left| {\psi \left( t \right)} \right\rangle $. In this work, the performance function 
${\cal {F}}$ used in the GRAPE optimization is explicitly defined to be the fidelity. The final control field $\hat{H}^{\rm{OPT}}(T)$ obtained after optimization effectively transforms the initial state into the target state within the theoretical and experimental limitations.

Within this framework, the Gradient Ascent Pulse Engineering (GRAPE) algorithm\cite{Khaneja2005} was employed to iteratively adjust the control parameters by computing the gradients of the fidelity with respect to the control parameters. To efficiently perform this optimization, the limited-memory Broyden-Fletcher-Goldfarb-Shanno algorithm with bound constraints (L-BFGS-B) was utilized within the GRAPE loop, which significantly accelerated convergence and ensured stable updates under experimental constraints.

In our optimal control implementation, the continuous phase optimization was reformulated as a constrained slice optimization problem to reduce experimental complexity. The total pulse duration $T$ was divided into $N$ non-uniform time slices with durations $\Delta {t_i}\left( {i = 1,2, \cdots ,N} \right)$, the constraint can be expressed as
\begin{equation}
\begin{aligned}
\sum\limits_{i = 1}^N {\Delta {t_i} = T} ,{\rm{     }}\left( {\Delta {t_{\min }} \le \Delta {t_i} \le \Delta {t_{\max }}} \right). \label{T}
\end{aligned}
\end{equation}
where $\Delta {t_{\min }}$ and $\Delta {t_{\max }} $ are the minimum and maximum slice durations. Within each slice duration, the laser phase $\phi_L(t)$  was assumed to remain constant and can be written as
\begin{equation}
\begin{aligned}
{\phi _L}\left( t \right) = {\phi _k},{\rm{    }}t \in \left[ {{t_{k - 1}},{t_k}} \right), \label{phi}
\end{aligned}
\end{equation}
where ${\phi _k} \in \left[ {0,2\pi } \right)$ represents the optimized phase value, and ${t_k} = \sum\limits_{i = 1}^k {\Delta {t_i}} $. This parameterization reduces the infinite-dimensional continuous optimization to $2N$-dimensional parameter space, consisting of $\left\{ {\Delta {t_i},{\phi _k}} \right\}$, thus preserving the ability to realize high-fidelity phase modulation while simultaneously reducing computational complexity.

In practice, the finite response time of experimental setups prevented instantaneous phase jumps between adjacent slices. To account for this effect, a linear response model was incorporated into the pulse design. The phase transition between two adjacent slices was modeled as a linear interpolation over a short duration ${\tau _{\rm{resp}}}$ at the beginning of each slice. The phase is redefined as
\begin{equation}
\begin{aligned}
\phi_{\exp}(t) = 
\begin{cases}
  \phi_{k-1} + \dfrac{(\phi_k - \phi_{k-1})}{\tau_{\rm{resp}}} (t - t_{k-1}), &t \in [t_{k-1}, t_{k-1} + \tau_{\rm{resp}}) \\
  \phi_k, &t \in [t_{k-1} + \tau_{\rm{resp}}, t_k)
\end{cases}\label{phi}
\end{aligned}
\end{equation}
This approach results in a piecewise linear phase profile that more accurately reflects the experimentally achievable waveform while maintaining the essential degrees of freedom for optimization. 

{To achieve transfer robustness, the optimization is performed over an ensemble of atomic configurations sampled from the Gaussian spatial and velocity distributions of the atomic cloud. Atoms located at different transverse positions experience different Rabi frequencies due to the Gaussian laser beam profile, while atoms with different longitudinal velocities acquire different Doppler-induced two-photon detunings. By uniformly sampling the distribution and evaluating the system dynamics for each sampled point $(r,v_z)$. This ensures that the resulting control pulse maintains high fidelity not only for the nominal system, but also remains robustness to the variations in detuning and coupling strength arising from the finite spatial and velocity spreads of the ensemble.} 

The central objective of mirror pulse optimization in atom interferometry was to realize high-fidelity quantum state transfer from the ground state ${\left| g \right\rangle _{r,{v_z}}}$ to the excited state ${\left| e \right\rangle _{r,{v_z}}}$. For a single atom, the fidelity was defined as
\begin{equation}
\begin{aligned}
{{\cal F}}_{\rm{single}}^{\rm{real}} = {\mathop{\rm Re}\nolimits} \left( {_{r,{v_z}}\left\langle e \right|\hat U\left( r,v_z;T,0 \right){{\left| g \right\rangle }_{r,{v_z}}}} \right),\\
{{\cal F}}_{\rm{single}}^{\rm{imag}}  = {\mathop{\rm Im}\nolimits} \left( {_{r,{v_z}}\left\langle e \right|\hat U\left( r,v_z;T,0 \right){{\left| g \right\rangle }_{r,{v_z}}}} \right). \label{F}
\end{aligned}
\end{equation}
where the propagator is given by
\begin{equation}
\begin{aligned}
\hat U(r,v_z;T,0) = {{\cal T}}\exp \left( { - \frac{i}{\hbar }\int\limits_0^{{T}} {\hat H\left(r,v_z, {t'} \right)dt'} } \right). \label{F}
\end{aligned}
\end{equation}

{In the following, we adopt the real-part definition ${\cal F_{\rm{single}}}={{\cal F}}_{\rm{single}}^{\rm{real}}$. This choice follows Ref. \cite{Saywell2018} and is essential for mirror pulses in AI. While maximizing the absolute value ensures population transfer, it does not constrain the final state phase across different velocity classes. Minimizing phase dispersion is crucial for preserving interferometric contrast, and therefore the real-part (or imaginary-part) fidelity is the appropriate metric for optimizing mirror pulses.}

{In our theoretical simulations, fidelity degradation of the mirror pulse arises from the assumed inhomogeneities of the modeled atomic ensemble. Because the spatial and velocity distributions are assumed to follow Gaussian profiles, different atoms experience different effective two-photon detunings determined and different Rabi frequencies. These distribution-induced variations constitute the primary source of inhomogeneity in the system.} To evaluate and enhance the pulse performance comprehensively, we constructed an ensemble averaged fidelity for the mirror pulse, which can be expressed as
\begin{equation}
\begin{aligned}
{{\cal F}_{\rm{ave}}} =\int\int{P_{sv}(r,v_z){\cal F_{\rm{single}}}drdv_z} . \label{FT}
\end{aligned}
\end{equation}
This fidelity takes into account the spatial and velocity probability distributions of atoms, thereby reflecting the robustness and effectiveness of the pulse under experimental conditions.

\section{RESULTS}\label{IV}
Using this model, the numerical simulation was performed with an atomic cloud characterized by a transverse Gaussian radius of $\sigma_r=3$ mm and a longitudinal velocity spread of $\sigma_{v_z}=5.4$ mm/s. {For clarity, the key optical and atomic parameters employed in the simulation are summarized in Table \ref{tab1}.} To generate a robust mirror pulse while reducing experimental complexity, a slice-based phase optimization strategy with a central symmetry constraint was employed. In this approach, the modulation sequence was composed of $N$ non-uniform time slices $\Delta t_i$. The central symmetry constraint on the phase profile effectively halved the number of independent optimization parameters, thereby streamlining the computational process. Within each slice duration, the Raman laser phase was kept constant, while the slice durations were allowed to adapt during the optimization.

\begin{table}[b]
\caption{\label{tab1}
Key optical and atomic parameters used in the numerical simulations, including the optical power, Raman detuning, Rabi frequency, laser beam waist, and the spatial and velocity characteristics of the atomic ensemble.}
\begin{ruledtabular}
\begin{tabular}{lc}
\textrm{Parameter} & \textrm{Value} \\
\colrule
Optical power & $P = 12.9~\mathrm{mW}$ \\
Single-photon detuning & $\Delta = 1.5\times 2\pi~\mathrm{GHz}$ \\
Rabi frequency & $\Omega_0 = 25 \times 2\pi~\mathrm{kHz}$ \\
Laser beam waist & $w = 11~\mathrm{mm}$ \\
Atomic cloud radius & $\sigma_r = 3~\mathrm{mm}$ \\
Transverse temperature & $T_{xy} = 3~\mu\mathrm{K}$ \\
Longitudinal temperature & $T_z = 300~\mathrm{nK}$ \\
\end{tabular}
\end{ruledtabular}
\end{table}

\begin{figure}[t]
\includegraphics[trim=20 5 30 5,width=\columnwidth]{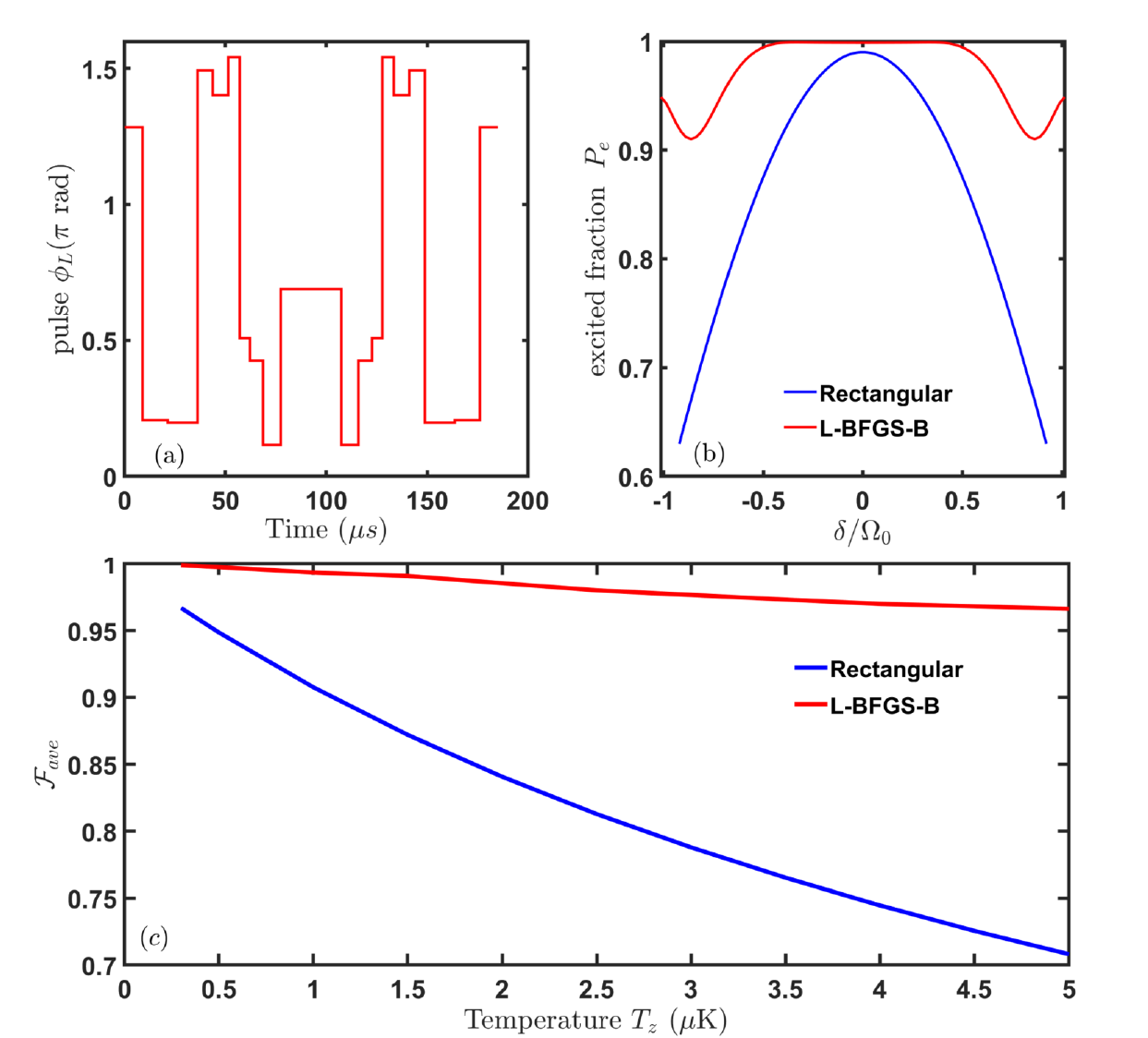}
\caption{\label{fig-phase} (a) Phase profile $\phi_L(t)$ for GRAPE pulse optimized to maximize $\cal{F}_{\rm{ave}}$ under the constraint of $N=20$ slices {($\tau_{\rm{resp}}=0~\mu$s)}. (b) Excited state population as a function of the two-photon detuning $\delta$, with the Rabi frequency fixed at $\Omega_0=2\pi \times 25$ kHz. Results for the optimized pulse phase (red curve) and the rectangular pulse phase (blue curve) are compared. (c) The ensemble averaged fidelity ${\cal{F}}_{ave}$ as a function of the longitudinal temperature $T_z$.}
\end{figure}

{Figure \ref{fig-phase}(a) shows the optimized phase profile $\phi_L(t)$ of the mirror pulse, corresponding to $\tau_{\rm{resp}}=0~\mu$s. }The optimization is performed within the parameter space of detuning $\delta\in\left[ { - {\Omega _{\rm{0}}},{\Omega _{\rm{0}}}} \right]$ and Rabi frequency $\Omega \in [ 0.5 \Omega_0, 1.5 \Omega_0 ]$, where $\Omega_0=2\pi \times 25$ kHz. The phase $\phi_L(t)$ was discretized into $N=20$ time slices and was adaptively adjusted (5 $\mu$s $\le \Delta {t_i} \le$ 15 $\mu$s) during the optimization. The resulting profile, indicated by the red curve, exhibits a step-like structure, reflecting the ability of the slice modulation to redistribute the phase dynamically across the pulse duration. Figure \ref{fig-phase}(b) illustrates the dependence of the excited state probability $P_e$ on the detuning for both the optimized pulse (red curve) and rectangular pulse (blue curve). Within the detuning range of $\left[ { -0.54\times{\Omega _{\rm{0}}},0.54\times{\Omega _{\rm{0}}}} \right]$, the optimized pulse demonstrates remarkable robustness with $P_e>0.99$ throughout, whereas the rectangular pulse shows a rapid decrease to ${P_e} \approx 0.85$ at $ \pm 0.54 \times {\Omega _{\rm{0}}}$, corresponding to an improvement of approximately $14\%$ in robustness. These results confirm that the slice phase modulation strategy effectively suppressed the adverse effect of the atomic cloud’s velocity distribution on the transition probability.

Figures. \ref{fig-phase}(a) and (b) present the performance at a fixed longitudinal temperature corresponding to the typical experimental conditions, it is also important to explore how the pulse performance depends on the atomic temperature. Figure \ref{fig-phase}(c) shows the dependence of the ensemble averaged fidelity $\mathcal{F}_{\rm ave}$ on the longitudinal temperature $T_z$, which is varied from $0.3~\mu$K to $5~\mu$K. As shown in Fig. \ref{fig-phase}(c), the fidelity associated with the rectangular pulse decreases rapidly with increasing temperature, dropping from above $0.9$ at low temperatures to approximately $0.71$ at $T_z=5~\mu$K. In contrast, the optimized pulse maintains a significantly higher fidelity across the entire temperature range, remaining above $0.96$ even at higher temperatures. These results demonstrate that the optimized pulse is highly robust against temperature induced Doppler broadening and remains effective under substantially relaxed experimental conditions.

\begin{figure}[t]
\includegraphics[trim=10 5 10 5,width=\columnwidth]{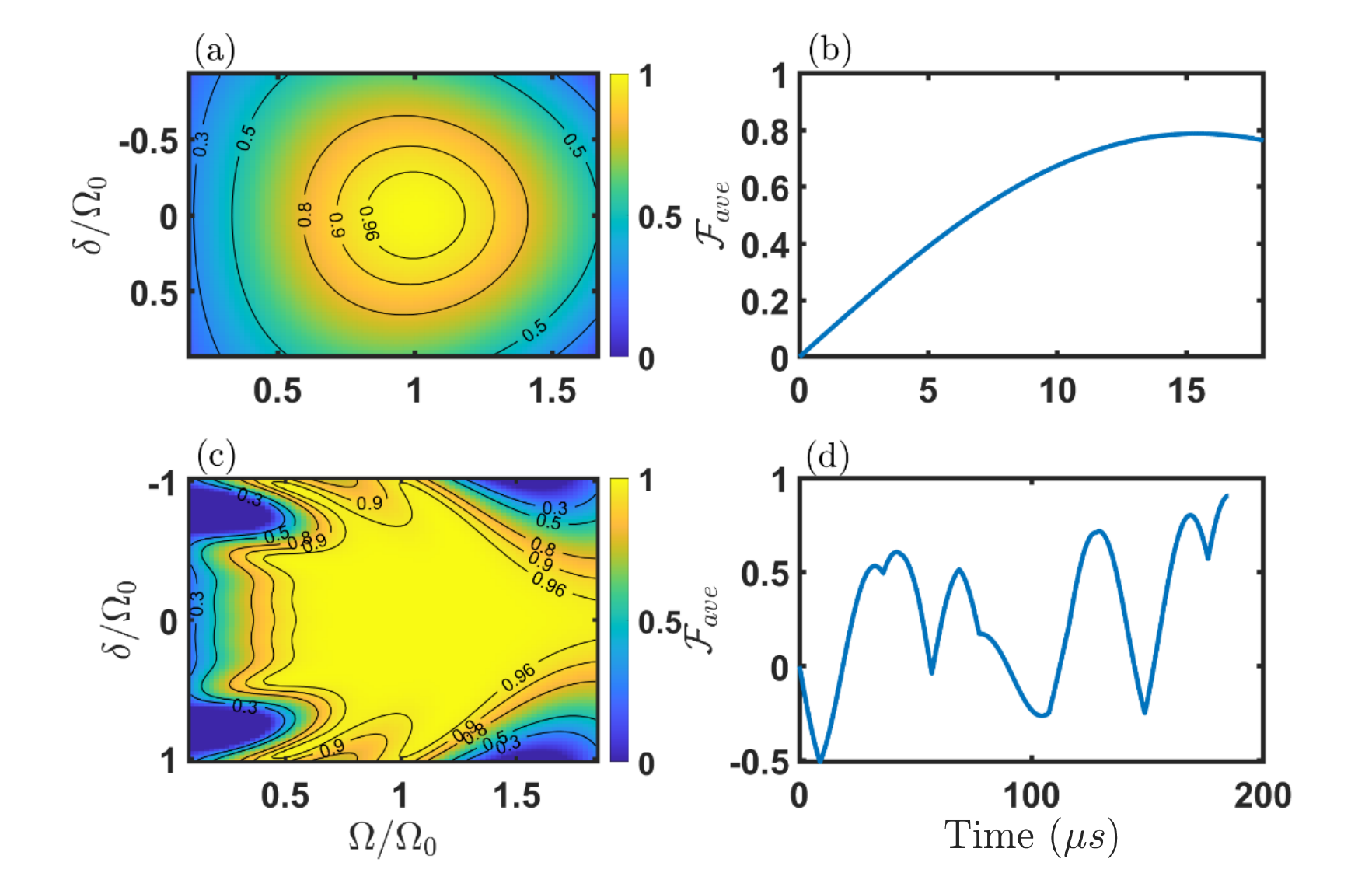}
\caption{\label{fig-fidelity} (a) and (c) show the excited state population in the parameter space of two photon detuning and Rabi frequency, for the rectangular phase pulse and optimized phase pulse, respectively. The contours are at 0.3, 0.5, 0.8, 0.9, and 0.96. (b) and (d) present the time evolution of the ${\cal{F}}_{ave}$ under the rectangular phase pulse and the optimized phase pulse, respectively. {All results are obtained assuming $\tau_{\rm{resp}}=0~\mu$s.}}
\end{figure}

The interferometer contrast was systematically investigated under variations in detuning $\delta$ and Rabi frequency $\Omega$ using the optimized phase profile from Fig. \ref{fig-phase}(a).  Figure \ref{fig-fidelity}(a,c) presents two-dimensional parameter scans comparing the excited state probability $P_e$ distributions between the rectangular pulses and the optimized sequence across variations in both detuning $\delta \in [-\Omega_0,\Omega_0]$ and coupling strength $\Omega \in [0.1\times\Omega_0,1.9\times\Omega_0]$, respectively. The optimized pulse exhibited strong robustness against laser intensity fluctuations with Rabi frequency varying over $\Omega \in [0.55\times\Omega_0,1.9\times\Omega_0]$, maintaining $P_e>0.96$. In contrast, the performance of the rectangular pulse degraded substantially, dropping to $P_e < 0.8$ in the same region, which corresponds to a relative efficiency loss of approximately $16\%$. These results confirmed that the slice-phase modulation strategy effectively suppressed the detrimental effects of both velocity inhomogeneity and intensity noise, thereby ensuring robust state transport.

 The temporal evolution of the ensemble averaged fidelity ${\cal{F}}_{\rm{ave}}(t)$ under both rectangular pulses and the optimized control scheme is shown in Fig. \ref{fig-fidelity}(b) and (d). For the optimized pulse, the fidelity exhibited persistent non-monotonic oscillations throughout the entire pulse duration, rather than a smooth increasing trend. This oscillatory behavior indicates that the system was driven along a nonadiabatic and dynamically rich trajectory. In contrast, the rectangular pulse led to a more gradual increase in fidelity, but its final performance remained inferior, the maximum ensemble averaged fidelity over the entire evolution reached only ${\cal{F}}_{ave}(T)\approx0.81$. At the terminal time, the optimized pulse sequence achieved ${\cal{F}}_{ave}(T)\approx0.93$, surpassing the rectangular pulse by a significant margin. These results highlight the robustness and efficiency of the slice phase modulation strategy in compensating for velocity dispersion and other inhomogeneities, despite the presence of non-monotonic dynamics during the evolution.

\begin{figure}[t]
\includegraphics[trim=30 5 40 5,width=\columnwidth]{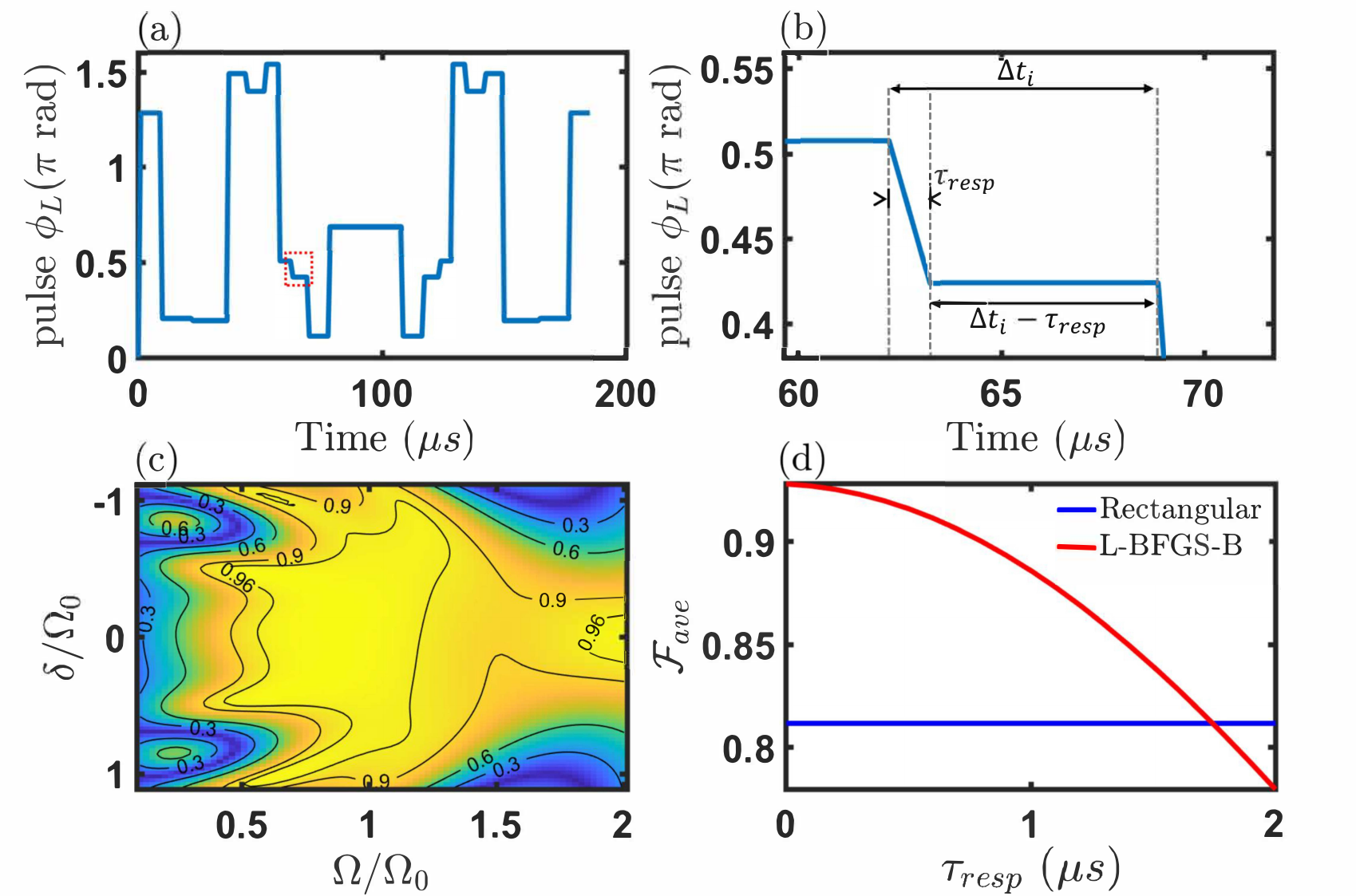}
\caption{\label{fig-tau} (a) Time evolution of the optimized phase pulse with a response time of $\tau_{\rm{resp}}=1~\mu$s. (b) Zoom of the red dashed box in panel (a), where the phase transition between two adjacent slices is treated as a linear transition with duration $\tau_{\rm{resp}}$. During the remaining interval $\Delta t_i-\tau_{\rm{resp}}$, the pulse phase is kept at the optimized value. (c) Excited state population of the system in the parameter space of Rabi frequency $\Omega$ and two-photon detuning $\delta$, taking into account the finite response time. (d) {Final ensemble-averaged fidelity ${\cal{F}}_{ave}(T)$ versus the response time, the optimized pulse (red) outperforms the rectangular pulse (blue) for $\tau_{\rm{resp}}\leq 1.6~\mu$s.}}
\end{figure}

To better reflect experimental constraints, the optimized control sequence was modified to incorporate a finite laser phase response time $\tau_{\rm{resp}}$. Figures~\ref{fig-tau}(a–c) show the results obtained for $\tau_{\rm{resp}}=1~\mu$s. The resulting phase profile is shown in Fig. \ref{fig-tau}(a), with a zoomed-in view of the boxed region in Fig. \ref{fig-tau}(b), where the linear phase transition between adjacent slices is explicitly illustrated. Figure \ref{fig-tau}(c) presents the robustness performance of the modified pulse across variations in laser detuning and Rabi frequency. The contour map indicates that, even with finite $\tau_{\rm{resp}}$, the optimized pulse maintained a high excitation probability across a broad parameter region in $\delta$ and $\Omega$. 

{In addition, Fig. \ref{fig-tau}(d) depicts the ${\cal{F}}_{\rm{ave}}(T)$ as a function of the response time $\tau_{\rm{resp}}$. As $\tau_{\rm{resp}}$ increases, the fidelity decreases monotonically due to the increasingly smoothed phase transitions that deviate from the optimized control sequence. Quantitatively, for $\tau_{\rm{resp}}\leq 1.6~\mu$s, the optimized pulse maintains a high fidelity above 0.81, which is significantly higher than that obtained with rectangular pulses, thereby confirming a clear performance advantage. However, when $\tau_{\rm{resp}} > 1.6~\mu $s, ${\cal{F}}_{\rm{ave}}(T)$ dropped below 0.81, and the optimized pulse no longer outperformed the rectangular case. These results demonstrate that the slice-phase modulation strategy retains robustness under realistic modulation constraints, while also providing practical guidance for the experimental choice of feasible response times.}

\begin{figure}[t]
\includegraphics[trim=10 5 10 5,width=\columnwidth]{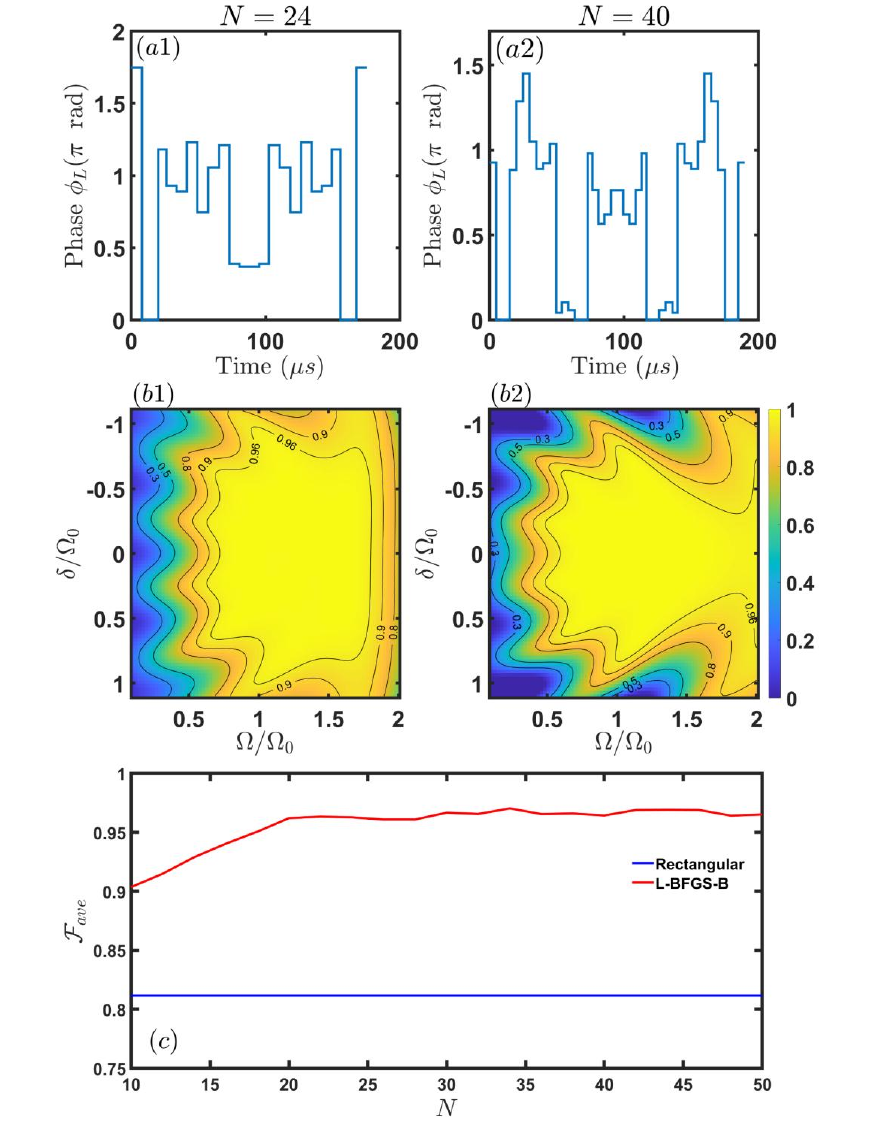}
\caption{\label{fig-N24-40} (a1, a2) Optimized phase profile of the control pulses with different numbers of time slices, $N=24$ and $N=40$. (b1, b2) Excited-state population of the system in the parameter space of $\Omega$ and  $\delta$ under the corresponding conditions. {(c) The ensemble-averaged fidelity ${\cal{F}}_{ave}$ as a function of the slice number $N$.}}
\end{figure}

In Raman-type AIs, the enhancement of transition efficiency through pulse optimization was found to be a general feature across different numbers of time slices. Figures \ref{fig-N24-40} (a1)-(b1) and (a2)-(b2) present two representative examples with $N=24$ and $N=40$, respectively. Figure \ref{fig-N24-40} (a1) shows the optimized phase profiles  for $N=24$, obtained by maximizing the ensemble averaged fidelity ${\cal{F}}_{\rm{ave}}(t)$. The corresponding robustness against detuning and coupling strength variations is presented in Fig \ref{fig-N24-40}(b1), where the excited-state probability remained above $P_e>0.96$ across the majority of the parameter space. For the case of $N=40$, similar improvements were observed, as illustrated in Figs. \ref{fig-N24-40}(a2)–(b2). In both cases, the final ensemble-averaged fidelities ${\cal{F}}_{\rm{ave}}(T)>0.9$, demonstrating that the proposed sliced-phase strategy consistently ensures robust state transfer under different discretizations. {To further quantify the influence of the slice number, we compute the final ensemble-averaged fidelity ${\cal{F}}_{\rm{ave}}(T)$ as a function of $N$, as shown in Fig. \ref{fig-N24-40}(c). The fidelity increases with the number of slices and enters a saturation regime for $N \ge 20$, where additional slices provide only marginal improvement. This behavior indicates that once sufficient control performance is reached, the sliced-phase pulse becomes insensitive to the precise value of $N$ and maintains consistently high performance. Thus, a moderate number of slices offers a favorable balance between robustness and control complexity. }

\section{Conclusion and Outlook}\label{V}
In this work, the mirror pulse were optimized with respect to the atomic velocity distribution and laser intensity variations to enhance the transition efficiency of AIs. A slice based phase optimization strategy was adopted, in which the detuning and Rabi frequency variations were incorporated into the ensemble distribution. Importantly, a finite phase response time was explicitly included in the optimization procedure. The results confirmed that, even under realistic modulation constraints, the optimized pulses maintained strong robustness against fluctuations in detuning and coupling strength, as well as robustness against variations in the longitudinal temperature, achieving a significant improvement over rectangular pulses.

This work presented an optimization of the mirror pulse in atom interferometer, while future studies could extend this approach to optimize beam splitter pulses and establish a comprehensive atom interferometer model, thereby holistically enhancing interferometer performance. Furthermore, while the current study focuses on static inhomogeneities, subsequent investigations may incorporate dynamic perturbations, such as coupling effects between atomic cloud position/velocity fluctuations and wavefront or magnetic field. Future work could incorporate decoherence effects within a density matrix formalism and explore machine learning-assisted real time optimization strategies to further strengthen robustness under complex environments. In addition, extending this approach to the control of multi-particle entangled states may open new directions for quantum-enhanced precision metrology.



\begin{acknowledgments}
This work was supported by the Key Researh and Development Program of China 
(Grant Nos. 2023YFC2907003 and 2022YFC3003802), National Natural Science Foundation of China (Grant Nos. U2341247, 12374464, 12204186 and 12274613). This work was also supported by the Open Fund of Wuhan Gravitation and Solid Earth Tides, National Observation and Research Station (Grant No. WHYWZ202410). Additionally, support was received from the Huazhong University of Science and Technology 2025 Provincial-Level College Students’ Innovation Training Program (Grant No. S202510487777).

\end{acknowledgments}

\section*{DATA AVAILABILITY}
The data that support the findings of this article are not publicly available. The data are available from the authors upon reasonable request.


\setcitestyle{showtitles}

\bibliographystyle{unsrt}

\bibliography{reference}


\end{document}